\documentclass[12pt]{iopart}
\usepackage{graphicx}
\usepackage{amssymb}
\usepackage{amssymb}
 \usepackage{latexsym}
 \usepackage{graphics}
\usepackage{graphicx}
 \usepackage{bm}
 \usepackage{color}
\newtheorem{lemma}{\sc \bf Lemma}[section]
\newtheorem{propos}{\sc \bf Proposition}[section]
\newtheorem{theor}{\sc \bf Theorem}[section]
\newtheorem{corr}{\sc \bf Corollary}[section]
\newtheorem{remark}{\sc \bf Remark}[section]
\newtheorem{definition}{\sc \bf Definition}[section]

\begin{document}
\title[Steady states in dual-cascade]{Steady states in dual-cascade wave turbulence}

\author{ V. N. Grebenev$^1$,
S. B. Medvedev$^2$, S. V. Nazarenko$^3$ and B. V. Semisalov$^{2,4}$
}
\address{$^1$  Faculty of Mathematics and Statistics, Ton Duc Thang University, Ho Chi Minh City, Vietnam}
\address{$^2$ Institute of Computational Technologies SD RAS,
Lavrentiev avenue 6, Novosibirsk 630090, Russia}
\address{$^3$ Insitute de Physique de Nice, Universite  C$\hat o$te D'Azur,  Ave. Joseph Vallot, Nice 06100,
France}
\address{$^4$ Novosibirsk State Univeristy,
Pirogova str. 2, Novosibirsk 630090, Russia}

\ead{ 
vngrebenev@gmail.com;
serbormed@gmail.com; Sergey.Nazarenko@unice.fr; vibis87@gmail.com}

\begin{abstract}
We study stationary solutions in the differential kinetic equation, which was
introduced in \cite{dpnz}
for description of a local dual cascade wave turbulence. We give a  full classification of single-cascade states in which
there is a finite flux of only one conserved quantity.
Analysis  of the steady-state spectrum is based on a phase-space analysis of orbits of the underlying dynamical system. The orbits of the dynamical system demonstrate the blow-up behaviour which corresponds to a ``sharp front" where the spectrum vanishes at a finite wave number.  The roles of the KZ and thermodynamic scaling as intermediate asymptotic, as well as of singular solutions, are discussed.
\end{abstract}
\maketitle

\section{Introduction}

\subsection{Wave turbulence kinetic equation and the differential approximation model}

Weak wave turbulence, a broadband system of random weakly interacting waves, is usually modeled by a kinetic equation
for the spectral wave action density, $n_{\bf k} \equiv n({\bf k , t})$, which describes how the excitations in the system are distributed among different wave vectors ${\bf k}$ at time $t$. The specific form of the wave kinetic equation depends on the linear and nonlinear properties of the medium, in particular the dispersion relation for the wave frequency, the leading-order of the resonant wave-wave interaction and the scaling properties of the nonlinear interaction term in PDE describing the medium dynamics.
For example,  a system dominated by four-wave ($2 \to 2$) interactions is described by the following kinetic equation~\cite{dpnz,zlf,Nazarenko},
\begin{eqnarray}\label{I1}
\frac{\partial n_{\bf k} }{\partial t} &=& \int |T_{{ {\bf k},{\bf k}_1,{\bf k}_2,{\bf k}_3}}|^2
\,  n_{\bf k}n_{\bf k_1}n_{\bf k_2}n_{\bf k_3}\left(\frac{1}{n_{\bf k}} + \frac{1}{n_{\bf k_1}} -
\frac{1}{n_{\bf k_2}} - \frac{1}{n_{\bf k_3}}\right) \times \\
&& \quad \delta({\bf k} + {\bf k}_1 - {\bf k}_2 - {\bf k}_3)
 \delta(\omega_{\bf k} + \omega_{{\bf k}_1} -  \omega_{{\bf k}_2} -  \omega_{{\bf k}_3})
\, d{\bf k}_1d {\bf k}_2d {\bf k}_3, \nonumber
\end{eqnarray}
where $T_{{ {\bf k},{\bf k}_1,{\bf k}_2,{\bf k}_3}}$ is an  interaction coefficient which depends on a particular wave system.
Assuming that  $T_{{ {\bf k},{\bf k}_1,{\bf k}_2,{\bf k}_3}}$ is strongly localised  at the values with
  ${ { {\bf k} \sim {\bf k}_1\sim{\bf k}_2\sim{\bf k}_3}}$,
equation~(\ref{I1}) can be approximated by a differential equation
\begin{equation}\label{I5}
\omega^{\frac d \alpha -1}
\frac{\partial n_\omega}{\partial t}=I\frac{\partial^2}{\partial\omega^2}\left(\omega^sn_\omega^4\frac{\partial^2}{\partial\omega^2}
\left(\frac{1}{n_\omega}\right)\right),
\end{equation}
where
\begin{equation}\label{I6}
n_\omega = n(k(\omega),t),
\end{equation}
\begin{equation}\label{I7}
s  = (2\gamma + 3d)/\alpha +2
\end{equation}
$d$ is the physical space dimension
and $I$ is a dimensional constant which depends on the particular physical system.
Here $\gamma$ is the degree of homogeneity of the interaction coefficient  defined via $
T_{{ \lambda {\bf k},\lambda {\bf k}_1, \lambda{\bf k}_2, \lambda{\bf k}_3}}
= \lambda^\gamma T_{{ {\bf k},{\bf k}_1,{\bf k}_2,{\bf k}_3}}$ where $\lambda$ is a constant.

\subsection{Physical applications}

Let us briefly list several important physical examples of four-wave systems.

{\it Wave turbulence in Bose-Einstein condensates and in Nonlinear Optics \cite{dpnz}.} In this case the underlying PDE is the Nonlinear Schr\"odinger (NLS, a.k.a Gross-Pitaevskii, GP) equation, and the resulting parameters are: $\alpha=2, \gamma = 0, d=2 $ or $3$. Respectively,
$s=5$ for 2D and $s=13/2$ for 3D systems.

{\it Wave turbulence on surface of deep water.} This is a classical and most celebrated example of wave turbulence \cite{zlf,Nazarenko,arfm}.
In this case $\alpha=1/2, \gamma = 3, d=2 $ so that $s=26$.

{\it Wave turbulence in self-gravitating dark matter.} Frequently used  model for self-gravitating dark matter is based on the so-called Schr\"odinger-Newton equations (SNE) \cite{SN}. Wave turbulence theory for such a system, including description within the differential approximation model, was recently developed in~\cite{SLN}. In this case the
parameters are: $\alpha=2, \gamma = -2, d=2 $ or $3$. Respectively,
$s=3$ for 2D and $s=9/2$ for 3D systems.

{\it Turbulence of gravitational waves in vacuum.} Wave turbulence theory for such a system was recently developed from the Einstein vacuum field equations in~\cite{GN}, and the differential model was further suggested in \cite{gnbt}.
In this case we have $\alpha=1, \gamma = 0$, so for  2D and 3D we have $s=8$ and $s=11$ respectively.

{\it Turbulence of Langmuir plasma waves and spin waves.} In this case we have $\alpha=2, \gamma = 2, d=3$. Respectively,
$s=17/2$.

\subsection{General properties of the differential approximation model}

 Differential approximation model~(\ref{I5}), is a rather drastic and not fully justified as a quantitative description for
most applications of weak turbulence. Nevertheless  retains many  qualitative properties of the
full kinetic equation~(\ref{I1}).
Differential kinetic equation~(\ref{I5}) has the conservation laws admitted within its original integro-differential form,
namely the conservation of the total energy,
\begin{equation}\label{I9}
E = \int \omega^{\frac d \alpha }
n_\omega \,
d\omega,
\end{equation}
and
the total number of particles,
\begin{equation}\label{I9}
N = \int \omega^{\frac d \alpha -1}
n_\omega \,
d\omega.
\end{equation}
It  also has the same the scaling and homogeneity as  the original kinetic equation. Consequently, the pure scaling
solutions of the kinetic equation, the equilibrium thermodynamic spectra and the non-equilibrium
Kolmogorov-Zakharov (KZ) cascade spectra, are also solutions of the differential kinetic equation (\ref{I5}). These solutions are of the form
$n(\omega) = c\cdot\omega^{-x}$ where $x$ takes one of the
following values:
\begin{eqnarray}\label{I10}
\hspace{-1.9cm}   x = 0  \, \hbox {(thermo N-equipartition)}, & \, x = 1 \, \hbox {(thermo E-equipartition)},\\
 \hspace{-1.9cm}  x = \frac{2\gamma + 3d}{3\alpha}
\, \hbox {(KZ inverse N-cascade)}, & \,
 x = \frac{2\gamma + 3d - \alpha}{3\alpha}
\, \hbox {(KZ direct E-cascade)}.
 \label{I11}
\end{eqnarray}

Quantity
\begin{equation}\label{5}
{\cal Q} = -\frac{\partial}{\partial\omega}{\cal K}
\end{equation}
has the meaning of the local flux of particles through $\omega$. We also define the local flux of energy~\cite{dpnz,CNP}:
\begin{equation}\label{6}
{\cal P} = {\cal K} -  \omega\frac{\partial{\cal K}}{\partial\omega},
\end{equation}
where
\begin{equation}\label{7}
{\cal K} =  I\,\omega^sn^4\frac{\partial^2}{\partial\omega^2}
\left(\frac{1}{n}\right).
\end{equation}

\subsection{Outline }

The differential kinetic equation (\ref{I5}) was  simulated numerically in
\cite{CNP} to find evolving spectra arising from decay of an initial data concentrated in a finite frequency range. It was found that the qualitative behaviour previously observed for a wave kinetic equation by Galtier et al in~\cite{GNNP} is also present in this model in the finite capacity case. In particular, in a direct cascade setup, these simulations show that the propagating front is sharp in the sense that
  $n(\omega) \equiv 0$ for $\omega > \omega_*(t)$  for $t < t_* < \infty$.
  The front position $\omega_*(t)$, as well as the evolution of the spectrum, were found numerically to possess properties
of  self-similarity of the second kind, so that  $\omega_*(t) \to \infty$ as $t  \to t_*$. In our future work, we will return to the study of the evolving self-similar solutions of the model  (\ref{I5}). In the present paper, we will concentrate on a detailed study and classifications of the stationary solutions of this model.

We are interested in an analysis full classification of the one-flux solutions equation~(\ref{10}) for for different values of
the problem parameter $s$. More exactly, we introduce a (KZ index) parameter $k = (s - 2)/3$ for the direct cascade solutions and $k = (s - 3)/3$ for the inverse cascade solutions, and show that there are three different types of behaviour in cases
$k > 1$, $k = 1$ and $k < 1$. Further, in each of these cases, different qualitative behaviour of spectra is shown to correspond to different sectors on the phase-plane of a respective 2D dynamical system, including KZ and thermodynamic spectra, as well as sharp cut-offs and blow up behaviour on the frequency boundaries of the ranges occupied by the spectra.

 The organisation of the article is as follows. In section~\ref{TTS}, after some rearrangements of independent and dependent variables of equation~(\ref{10}), we present the equation together with the dynamical systems both for the direct and indirect cascade solutions which make it suitable for finding the stationary solutions. Sections~\ref{S}--\ref{B} are devoted to an analysis of the dynamical systems based on phase portraits for different values of $k$. Namely, we consider the following range of parameters $k > 1$, $k = 1$
and $0 < k < 1$.  The full classification is given in terms of sets of the qualitatively different orbits. A summary and discussion of results is given in section~\ref{C}.

\section{Stationary solutions}\label{TTS}

In this paper, we study the stationary solutions of the differential kinetic equation~(\ref{I5}), i.e. solutions to the following stationary version of this equation:
\begin{equation}\label{11A}
I \frac{\partial^2}{\partial\omega^2}\left(\omega^sn^4\frac{\partial^2}{\partial\omega^2}
\left(\frac{1}{n}\right)\right) =0.
\end{equation}
Integrating this equation twice, we obtain
\begin{equation}\label{10}
I\, \omega^sn^4\frac{\partial^2}{\partial\omega^2}
\left(\frac{1}{n}\right) = {\cal P} - {\cal Q} \, \omega,
\end{equation}
where ${\cal P}$ and ${\cal Q}$ are arbitrary constants having the meanings of the energy and the particle fluxes
given by formulae  (\ref{6})  and  (\ref{5}).
Therefore the investigation of steady states of equation~(\ref{11A}) is reduced to the  study of  equation~(\ref{10}) for different choices of the constant fluxes ${\cal P}$ and ${\cal Q}$. In the present work, we will restrict ourselves to the cases where either
${\cal P}=0$ or ${\cal Q}=0$. Moreover, having in mind physically realistic situations only, with $n \ge 0$, we must take  ${\cal P} \ge 0$ and ${\cal Q}\le 0$, which a version of the Fj{\o}rtoft-Kraichnan dual-cascade statement~\cite{fjortoft,kraichnan,NazarenkoNewell}.~\footnote{The Fj{\o}rtoft-Kraichnan dual-cascade statement is that, in turbulent systems with two positive quadratic invariants, one of the invariants cascades downscale and the other--upscale from the forcing scale. In general, this statement is asymptotic: it applies to situations where the forcing and the dissipation regions are separated by large inertial intervals. In the differential models case, the inertial ranges do not have to be large: either ${\cal P}$ or ${\cal Q}$ vanish right outside of the forcing  range, provided that there is only one such range. Mixed state with simultaneous ${\cal P},\, {\cal Q}\ne0$  can occur only in between of two (or more) forcing regions.}
The solutions with ${\cal P}=0$, ${\cal Q}\ge0$  (or ${\cal P}\le0$, ${\cal Q}=0$) can be obtained from the solutions considered in the present paper using the symmetry in the equation (\ref{10}) with respect to
${\cal P}\to -{\cal P},\, {\cal Q}\to -{\cal Q},\, n\to -n$. Thus, in such solutions $n \le 0$ which makes them unphysical.

Note that in the cases when one of the fluxes is zero, equation~(\ref{10}) is of the Emden-Fowler type~i.e.
\begin{equation}\label{10a}
\frac{d^2y}{dx^2} = \pm x^{\beta}|y|^{\sigma}{\rm sgn}\,y
\end{equation}
after a change of  variable  $y = 1/n$ and identifying $\omega = x$. The asymptotic behavior of $y(x)$ as $x \to \pm\infty$
was extensively studied in many works, see e.g. \cite{Bellman}. This equation is integrable under $\sigma + \beta = 0$ and the solution is given in terms of the hypergeometric function, see for details~\cite{Mehta}. The generalized form of~(\ref{10a}) (which extends on the equation~(\ref{10}) with arbitrary fluxes) is
\begin{equation}\label{10b}
\frac{d^ny}{dx^2} = p(x)|y|^{\sigma}{\rm sgn}\,y,
\end{equation}
and the asymptotic analysis of solutions with respect to the growth properties of $p(x)$ as $x \to \pm \infty$ were presented in~\cite{Samovol}.


To study the one-flux steady states for equation~(\ref{10}), we perform the following transformations both for independent and dependent variables,
\begin{equation}\label{11}
\tau=\ln\omega,\quad u=\frac{1}{\omega^{k}n},\quad k=\frac{s-2}{3}
\end{equation}
for the cases with ${\cal Q} = 0, {\cal P}\ge 0 $ (direct cascade). Respectively, for ${\cal P} = 0, {\cal Q}\le 0 $ (inverse cascade) we transform as
\begin{equation}\label{11a}
\tau=\ln\omega,\quad u=\frac{1}{\omega^{k}n},\quad k=\frac{s-3}{3}.
\end{equation}
As a result, we get an autonomous semi-linear ODE  for both one-flux situations (${\cal Q} = 0, {\cal P}\ge 0 $ and ${\cal P} = 0, {\cal Q}\le 0 $) of the form
\begin{equation}\label{S2}
u_{\tau\tau} + (2k-1)u_{\tau} +k(k-1)u = Bu^4,
\end{equation}
where $B$ is always a non-negative number, $B = {\cal P}{ I}^{-1}$ for ${\cal Q} = 0$ and $B = - {\cal Q}{ I}^{-1}$ for
${\cal P} = 0$.

For equation~(\ref{S2}) with $B > 0$, we are interested in full classification of solutions with different behaviours
of $u(\tau)$ (and, respectively, the spectrum $n(\omega)$) both for finite and infinite ranges of $\tau$ or $\omega$. Settings of the boundary value problems and analysis of existence of their solutions are based on a phase--space analysis of orbits of the underlying dynamical system.


It will be convenient to write down equation~(\ref{10})  with $B = 1$,
\begin{equation}\label{S6}
u_{\tau\tau} + (2k-1)u_{\tau} +k(k-1)u - u^4 = 0,
\end{equation}
taking into account the fact that equation~(\ref{S2}) admits a scaling group of transformations of independent
variable $\tau$ and dependent variable $u$.

The easiest way to investigate solvability of boundary value problems for equation~(\ref{S6}) is to consider the phase space plot of the respective autonomous dynamical system associated with equation~(\ref{S6}):
\begin{eqnarray}
&&\frac{du}{d\tau} = v, \label{S7}\\
&&\frac{dv}{d\tau} = -(2k-1)v-k(k-1)u+u^4 \label{S8}.
\end{eqnarray}
For equilibria we have either $u = v = 0$  or
$u =\left(k(k-1)\right)^{\frac{1}{3}}$, $v = 0$. Therefore, we always have fixed point $P1 = (0, 0)$,
and sometimes also fixed point $P2 = (\left(k(k-1)\right)^{\frac{1}{3}},0)$. The latter exists only
for $k > 1$ and $k < 0$ since from the physics $u$  must be a non-negative function. The linearised version of the dynamical
system near the fixed point $P1 = (0, 0)$ reads
\begin{equation}\label{S9}
\frac{d}{d\tau}\left(\begin{array}{c}u\\v\end{array}\right)=
\left(\begin{array}{cc}0&1\\-k(k-1)&-(2k-1)\end{array}\right)\left(\begin{array}{c}u\\v\end{array}\right)
\end{equation}
with  eigenvalues
$\lambda_1 = 1 -k$ and $\lambda_2 = -k$ and eigenvectors $\xi_1 = (1,1-k) \equiv (1,\lambda_1)$ and $\xi_2 = (1,- k) \equiv (1,\lambda_2)$.
Correspondingly, near  $P2$ we have the following linearised system,
\begin{equation}\label{S10}
\frac{d}{d\tau}\left(\begin{array}{c}u\\v\end{array}\right)=\left(\begin{array}{cc}0&1\\3k(k-1)&-(2k-1)\end{array}\right)
\left(\begin{array}{c}u\\v\end{array}\right).
\end{equation}
The eigenvalues are given by
$\lambda_1 = -k+\frac{1}{2}+\frac{1}{2}\sqrt{16k(k-1)+1}$, $\lambda_2 = -k+\frac{1}{2}-\frac{1}{2}\sqrt{16k(k-1)+1}$
and eigenvectors $\xi_1 = (1,\lambda_1)$ and $\xi_2 = (1,\lambda_2)$.


\section{Phase space analysis for $k > 1$}\label{S}

Physical examples with $k > 1$ include:
3D NLS direct cascade ($k=3/2$), 3D NLS inverse cascade ($k=7/6$), deep water gravity wave direct cascade ($k=8$) and inverse cascade ($k=23/3$), gravitational waves direct cascade ($k=2$ in 2D and $k=3$ in 3D) and inverse cascade
($k=5/3$ in 2D and $k=8/3$ in 3D), Langmuir and spin waves direct cascade ($k=13/6$) and inverse cascade
($k=11/6$).


To analyse the behavior of orbits, we
supplement equation~(\ref{S6}) by the initial conditions
\begin{equation}\label{S6C}
u(0) = u_0 > 0, \qquad \left.\frac{d}{d\tau}u\right|_{\tau=0} = 0.
\end{equation}
These initial conditions mean that we study the orbits of the dynamical system which start on the $u$-axis of the phase plane $(u,v)$. Notice that the direction of velocity $(du/dt,dv/dt)$ on the $u$-axis is directed into the fourth quadrant for $u < \left(k(k-1)\right)^{\frac{1}{3}}$~i.e. to the left from the stable-node fixed point $P1$ (only $u>0$ half-plane is physically relevant),  and this vector field is directed out side of the fourth quadrant when $u > \left(k(k-1)\right)^{\frac{1}{3}}$~i.e. to the right from the equilibria $P1$. Correspondingly, the vector field $(du/dt,dv/dt)$ is directed outside of the fourth quadrant on the $v$-axes for $v < 0$ and inside of the first quadrant on the
$v$-axes for $v > 0$.
Notice that equation~(\ref{S6}) has an exact positive solution $(u,v)=(u_{P2},0)$ with $u_{P2} = \left(k(k-1)\right)^{\frac{1}{3}}$ which corresponds to the saddle point $P2$ on the phase plane and also exists
the trivial solution $(u,v)=(u_{P1},v_{P1}) = (0,0)$ which corresponds to the stable node $P1$.
Solution $(u,v)=(u_{P2},0)$ is the famous KZ spectrum with $n \sim \omega^{-k}$.

We begin with a preliminarily analysis of the behaviour of orbits of the dynamical system~(\ref{S7}), (\ref{S8}) which start to left  from the fixed point $P2$.
\begin{lemma}\label{L1}
The orbits of the dynamical system~(\ref{S7}),~(\ref{S8}) intersecting the $u$-axis with $u \leq \left(k(k-1)\right)^{\frac{1}{3}}$ go to the fixed point $P1$.
\end{lemma}
This assertion is natural considering the fact that the system~(\ref{S7}),~(\ref{S8}) can be interpreted as a motion of Newtonian particle is a system with potential $\Phi(u)= k(k-1) u^2/2 - u^5/5$ (with a minimum at $u=0$ and a maximum at $u=u_{P2}$) and a positive friction.
To prove this Lemma, we show that for equation~(\ref{S6}) the maximum (minimum) principle  holds in the following form.
\begin{propos}\label{P1}
For any  $\tau_{\min} < \infty$,  solutions $u(\tau)$ of the Cauchy problem~(\ref{S6}),~(\ref{S6C}) cannot achieve a non-negative minimum at $\tau = \tau_{\min}$
with $0 \leq u_{\min} = u(\tau_{\min})  < \left(k(k-1)\right)^{\frac{1}{3}}$.
\end{propos}
If $u_{\min} > 0$ then the assertion of this proposition immediately follows from a simple analysis of signs of  terms of equation~(\ref{S6}). For  $u_{\min} = 0$ and since this is a minimum, $\left.u_{\tau}\right|_{\tau = \tau_{\min}} = 0$, we have $(u_{\min}, v_{\min}) = (0,0)$. However, since this is a fixed point, it cannot be reached at a finite ``time", $\tau_{\min} < \infty$.

\begin{propos}\label{C1}
Monotone decreasing solution~$u(\tau)$ cannot asymptote to a constant $c$ as $\tau \to \infty$ excepting the values of $c$ equals $0$ or
$c = \left(k(k-1)\right)^{\frac{1}{3}}$.
\end{propos}
The proof of this assertion is evident.
\begin{propos}\label{P2}
Derivative $u_{\tau}$ is a uniformly bounded function on the set $\{\tau\,:\, u(\tau) \geq 0 , u(0) \leq \left(k(k-1)\right)^{\frac{1}{3}} \}$.
\end{propos}
In order to prove it, we multiply equation~(\ref{S6}) by $u_{\tau}$ and integrate over $[0,\tau]$, $\tau \leq \tau^{**}$ with $u(\tau^{**}) =0$.
As a result, we get the following integral (energy balance) identity,
\begin{equation}\label{S11}
\frac{1}{2}u_{\tau}(\tau)^2 + (2k-1)\int_0^\tau u_s^2ds  + \Phi(u(\tau)) = \Phi(u_0),
\end{equation}
where $\Phi(u) = \frac{k(k-1)u^2}{2} - \frac{1}{5}u^5$. We have $\Phi(u) > 0$ for $0 < u < \left(\frac{5}{2}k(k-1)\right)^{\frac{1}{3}}$
and $\Phi(u) < 0$ for $u > \left(\frac{5}{2}k(k-1)\right)^{\frac{1}{3}}$.  $\Phi(u)$ achieves a positive maximum at $u = \left(k(k-1)\right)^{\frac{1}{3}}$. It follows from Proposition~\ref{P1} that $\Phi(u(\tau)) \geq  0$,
since $u_0 \leq \left(k(k-1)\right)^{\frac{1}{3}}$. Therefore $u_{\tau}^2(\tau)$ is an uniformly bounded function for $\tau \in [0,\tau^{**}]$.

To proceed with the proving Lemma~\ref{L1}, we assume the contrary i.e. that there exists an orbit $U_{u_0}$ which does not achieve the fixed point $P1$. By Proposition~\ref{P1}, it is evident that $U_{u_0}$ has to approach the $v$-axis, and due to Proposition~\ref{P2} the orbit has to intersect the $v$-axis at a finite point with $v<0$. Further, with an inverse time direction $\hat\tau = -\tau$, take an orbit  $U_{\xi_1}$ which goes out of $P1$ along the eigenvector $\xi_1$: then $U_{\xi_1}$ is above  the orbit $U_{u_0}$ on the phase plane. Considering the fact that in the fourth quadrant the vector field at infinity is pointing into the fourth quadrant,
by the Poincar\'e-Bendixon theorem $U_{\xi_1}$ always intersects the $u$-axis. In terms of solutions of the Cauchy problem ~(\ref{S6}),~(\ref{S6C}), the orbit $U_{\xi_1}$ ($U_{u_0}$) corresponds to $u^{\xi_1}$ ($u^{u_0}$), a solution of~(\ref{S6}),~(\ref{S6C}). $u^{\xi_1}(t)$ is a  positive  function which monotonously decreases to zero as $\tau \to \infty$. The latter property follows from Proposition~\ref{P1}.
 In fact, $u^{\xi_1}(\tau)$ is a sub-solution for $u^{u_0}(\tau)$. It means that $u^{u_0}(\tau)$ is a monotonously decreasing function such that $u^{u_0}(\tau) \to 0$ as $\tau \to \infty$. Hence  $U_{u_0}$ intersects $v$-axis  at $P1$ only, which proves Lemma~\ref{L1}.

Consider a family $O_I^a$, the orbits  $U_{u_0}$ with $0 < u_0 < \left(k(k-1)\right)^{\frac{1}{3}}$. This family depends continuously
on $u_0$ that follows from the continuous dependence of solutions $u(\tau;u_0)$ of the Cauchy problem~(\ref{S6}),~(\ref{S6C}) with respect the initial data. Moreover, the first-order derivatives of $u(\tau;u_0)$ are uniformly bounded functions with respect to $u_0$.
Hence there exist the limit of $U_{u_0}$ as $u_0$ goes to $\left(k(k-1)\right)^{\frac{1}{3}}$ and this is exactly the heteroclinic orbit $H$ which connects the fixed points $P1$ and $P2$.

For analysing the other orbits from the fourth quadrant of the phase space $(u,v)$, we will use the fact that the vector field  $(du/d\tau,dv/d\tau)$  is directed out of the fourth quadrant on the $v$-axis. The slope of this vector on the $v$-axis is always constant i.~e. $du/dv = - 1/(2k - 1)$. We denote the corresponding orbits  by $O_{II}$.
With the inverse time $\hat \tau$, these orbits cannot leave the fourth quadrant of the phase plane due to the fact that on the $u$-axis for $u > \left(k(k-1)\right)^{\frac{1}{3}}$, the vector field $(du/d\hat t,dv/d \hat t)$   is directed inside  the fourth quadrant: we call these orbits $O_{III}^a$. Therefore there exists an orbit $S_{II,III}$ which separates the families $O_{II}$ and $O_{III}^a$. This is exactly a separatrix passing through the saddle point $P2$. By the Poincar\'e-Bendixon theorem it
follows that the separatrix $S_{II,III}$ goes to infinity never approaching  the $u$- or $v$-axes.

To complete analysing the phase-space plot  of the dynamical system~(\ref{S7}),(\ref{S8}), we turn our attention to the orbits in the first quadrant of the phase plane. We will use  the fact that on the $v$-axis the vector field  $(du/d\tau,dv/d\tau)$   is directed into  the first quadrant. With this and using the information about the behavior of the vector field $(du/d\hat\tau,dv/d \hat\tau)$  on the $u$-axis, we get that there exists an orbit $S_{IV,I}$ intersecting $P2$  separating the orbits which cross and do not cross the $u$-axis respectively. In the inverse time direction, $\hat \tau = -\tau$, the separatrix $S_{IV,I}$ has to intersect the $v$-axis at a finite point. The proof follows from analysing again the integral identity~(\ref{S11}) which also holds for  $S_{IV,I}$. Therefore there exist orbits intersecting the $v$-axis, $O^b_{I}$, which enter the first quadrant, in the positive time direction, and then intersect the $u$-axis with $u \leq \left(k(k-1)\right)^{\frac{1}{3}}$.
Since the fixed point $P2$ is a saddle point then there exists a separatrix $S_{III,IV}$ which expands into the first quadrant. $S_{III,IV}$ separates the orbits with different behaviours, $O_{III}^b$  and $O_{IV}$.
\begin{lemma}\label{L2}
The orbits of the dynamical system~(\ref{S7}),~(\ref{S8}) intersecting the $u$-axis with $u \geq \left(k(k-1)\right)^{\frac{1}{3}}$ go to infinity never asymptoting  the line $u = const$~i.e. these orbits extend indefinitely.
\end{lemma}
The proof follows immediately from the integral identity
\begin{equation}\label{S12}
\frac{1}{2}u_{\tau}(\tau)^2 + (2k-1)\int_0^{\tau} u_s^2ds  + \Phi(u(\tau)) = \Phi(u_0).
\end{equation}
Indeed, assuming the contrary we get that $\Phi(u(\tau))$ is finite and
therefore $u_{\tau}(\tau) < \infty$ as $\tau \to \infty$
that guarantees that the orbit under consideration does not approach the line  $u = const$ with time.

\begin{figure}[h]
\center{\includegraphics[width=8cm]{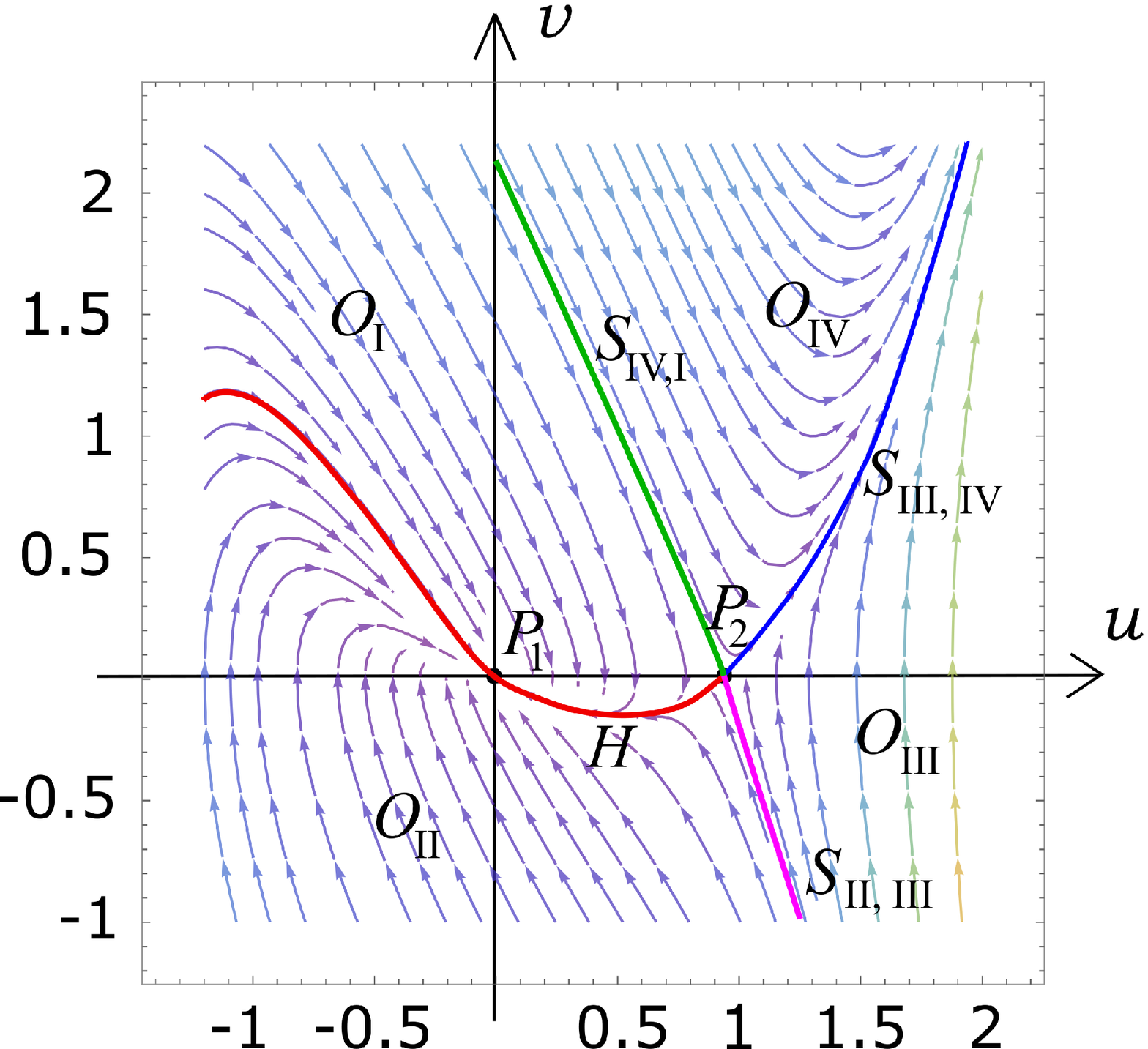}}.
\caption{Phase portrait of the dynamical system for $k = 3/2$.
The separatrixes are shown by different color lines. }
\label{portrait1}
\end{figure}

We will now put together the classification of the orbits, see~Figure~\ref{portrait1}.

\begin{theor}\label{T1}
$H\cup S_{II,III}\cup S_{III,IV}\cup S_{IV,I}$ divide the phase plane $(u,v)$ $(0 < u < \infty,  -\infty < g < \infty)$ into the parts $O_I = O_{I}^a\cup O_{I}^b$, $O_{II}$, $O_{III}= Q_{III}^a\cup Q_{III}^b$ and $O_{IV}$ with the different behaviours of orbits. In bounded $O_I$,  the orbits go to the fixed $P1$ and always intersecting $u$-axis with $u \leq \left(k(k-1)\right)^{\frac{1}{3}}$. In part $O_{II}$ orbits goes from infinity of the fourth quadrant intersecting the $v$-axis (never intersecting $u$-axis).  In part $O_{III}$  orbits emerge with $(u,v) \to (\infty, -\infty)$, go down along $S_{III,IV}$ separatrix, intersecting $u$-axis with $u \geq \left(k(k-1)\right)^{\frac{1}{3}}$, and then turn up  to extending into the first quadrant never intersecting $v$-axis. The  $O_{IV}$ orbits enter into the first quadrant intersecting the $v$-axis above   $S_{IV,I}$ separatrix and then extend with $(u,v) \to (\infty, \infty)$ never intersecting the $u$-axis.
\end{theor}

Now we will consider how the orbits classified in Theorem~\ref{T1}
could be linked to the classification of solutions of  the original equation.
First,  we note that on the $u$-axis
 the profiles $u(\tau)$ reach maxima, which corresponds to minima of the ``compensated" spectrum
 $\omega^k n(\omega)$.
 In the other words, on the $u$-axis the spectrum  $ n(\omega)$ changes from being steeper than KZ (above the axis) to being shallower than KZ.

We proceed by considering spectra corresponding to
the separatrix $S_{III,IV}$ and show that $u(\tau)$, which corresponds to $S_{III,IV}$, blows up at
a finite time $\tau_*$ i.e. $u(\tau) \to \infty$ as $\tau \to \tau_*$. We know that $S_{III,IV}$ goes to infinity in both $u$ and $v$. Further, for $u \gg 1$ and $v \gg 1$, the dynamics
of this separatrix in the leading order is governed by
\begin{eqnarray}
&&\frac{du}{d\tau} = v, \label{S7a}\\
&&\frac{dv}{d\tau} = u^4 \label{S7b}
\end{eqnarray}
since $v \sim u^{5/2} \ll u^4$ for $u \gg 1$. The system $(\ref{S7a},\ref{S7b})$ is a Hamiltonian type with the Hamilton function
$H = v^2/2 - u^5/5$. Integrating in the asymptotic limit $u^5/5 \gg H$, we get
\begin{equation}\label{S12a}
u(\tau) = (2/3)^{2/3}(5/2)^{1/3}(\tau_* - \tau)^{-2/3}, \quad \tau < \tau_* < \infty.
\end{equation}
Respectively, for the spectrum $n(\omega)$, we have
\begin{equation}\label{S12b}
\!\!\!\! \!\!\!\! \!\!\!\! \!\!\!\! \!\!\!\! \!\!\!\! \!\!\!\! \!\!\!\!  \!\!
n(\omega) = \omega^{-2/3}(3/2)^{2/3}(2/5)^{1/3}\ln^{2/3}(\omega_*/\omega)
\approx \omega_{*}^{-4/3}(4/15)^{1/3}  (\omega_* - \omega)^{2/3}
,
\; \omega < \omega_* < \infty,
\end{equation}
i.e. a sharp front on the right.

The behaviour of $u(\tau)$ which corresponds to the separatrix $S_{II,III}$ is analysed similarity. It can be done in terms of
the reverse time $\hat\tau$. As a result, we get  the asymptotic solution
\begin{equation}\label{S12b}
u(\hat\tau) = (2/3)^{2/3}(5/2)^{1/3}(\hat\tau^{**} - \hat\tau)^{-2/3}, \quad  \hat\tau < \hat\tau^{**} < \infty
\end{equation}
and in the original variables $\omega$, $n$ we have
\begin{equation}\label{S12b}
n(\omega)
\approx \omega_{**}^{-4/3}(4/15)^{1/3}   (\omega - \omega_{**})^{2/3}
, \quad \omega_{**} < \omega < \infty,
\end{equation}
i.e. a sharp front on the left.

Similarly, we consider the orbits located between the separatrices $S_{III,IV}$ and $S_{II,III}$ i.e. the ones belonging to
$O_{III}$. It follows from above that  the spectrum $n(\omega)$ corresponding to  these orbits
realises profiles with finite supports (i.e. with sharp fronts on both right and left).

Consider behaviour near the $v$-axis. In the leading order, the dynamical system~(\ref{S7},\ref{S8})
is reduced to
\begin{eqnarray}
&&\frac{du}{d\tau} = v, \label{S8a}\\
&&\frac{dv}{d\tau} = -(2k - 1)v \label{S8b}
\end{eqnarray}
for $u \ll v$. Solving this system, we have
\begin{equation}\label{S8c}
\!\!\!\! \!\!\!\! \!\!\!\! \!\!\!\! \!\!\!\! \!\!\!\! \!\!\!\! \!\!\!\!
u = \frac{v_0}{2k - 1}\left[e^{-(2k - 1)\tau_0}-e^{-(2k - 1)\tau} \right], \;\; n = \frac{2k - 1}{v_0(\omega_0^{1 - 2k} - \omega^{1 - 2k})\omega^k} \approx
 \frac{\omega_0^{ k}}{v_0 }\frac{1}{(\omega - \omega_0)}
.
\end{equation}
We see that spectrum $n(\omega)$ blows up at a finite point $\omega_0$: at its left boundary for
 for $v > 0$ ($O_{I}$ orbits) and at its right boundary for  for $v <0$ ($O_{II}$ orbits).

To complete, we consider orbits  near  the equilibrium point $P1$,
\begin{equation}\label{S8d}
u =   c_1 e^{(-k + 1)\tau} + c_2 e^{-k\tau},
\end{equation}
where $c_1, c_2= const$. This corresponds to
a thermodynamic Rayleigh-Jeans spectrum
\begin{equation}\label{S8e}
n = \frac{T}{\mu + \omega},
\end{equation}
where $T$ and $\mu$ are constants  having meanings of a temperature and a chemical potential respectively.
In fact, since the close vicinity of $P1$ corresponds to $\tau, \omega \to \infty$, we have
$n \approx T/\omega$, except for a single orbit (entering $P1$ along the second eigenvector direction)
for which $n=$const.
Positivity of $n$ for $\omega \to \infty$ dictates that $T>0$.

Easy to see that spectra (\ref{S8c}) are also thermodynamic--they can be written in the form
(\ref{S8e}) after a suitable choice of  constants $T$ and $\mu$.
In particular spectra with $v_0>0$ correspond to $T>0, \mu<0$ and spectra with $v_0<0$ -- to
 $T<0, \mu<0$.
 Note that the solution corresponding to the separatrix $H$ has $\mu =0$.

For the separatrix $S_{IV,I}$, the corresponding solution $u(\tau)$ is a function monotone increasing  from zero to
$u = (k(k-1))^{1/3}$, i.e. to the equilibrium $P2$ which corresponds to the KZ spectrum $n = (k(k-1))^{-1/3}\omega^{-k}$.

\begin{figure}[htbp]
\begin{center}
\includegraphics[width=12cm]{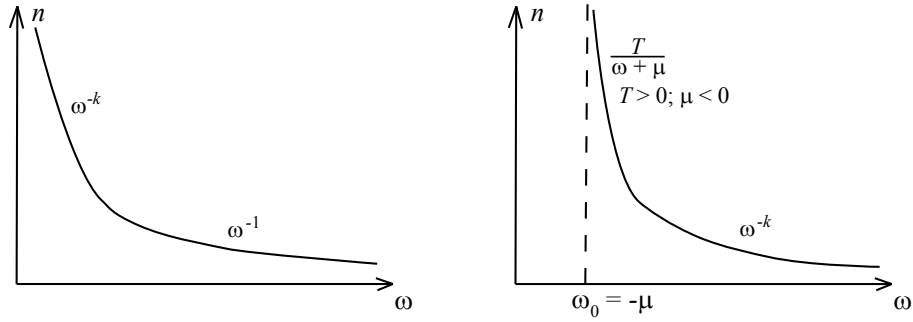}
\end{center}
\caption{ Sketches of spectra in case $k >1$.
{\bf Left:} separatrix $H$.
{\bf Right:} separatrix $S_{IV,I}$.
}
\label{portrait2S}
\end{figure}

\begin{figure}[htbp]
\begin{center}
\includegraphics[width=12cm]{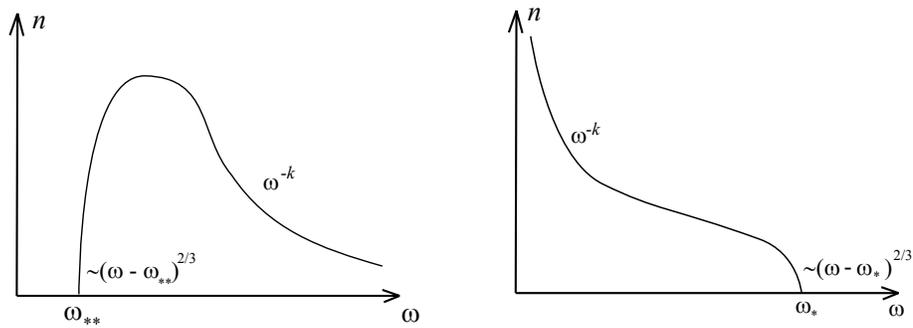}
\end{center}
\caption{ Sketches of spectra in case $k >1$.
{\bf Left:}  separatrix $S_{II,III}$.
{\bf Right:}  separatrix $S_{III,IV}$.
\
}
\label{portrait3S}
\end{figure}

The behavior of the spectra corresponding to the separatrices $H$, $S_{IV,I}$, $S_{II,III}$ and $S_{III,IV}$
are presented on Figure~\ref{portrait2S} and Figure~\ref{portrait3S}.

\begin{figure}[htbp]
\begin{center}
\includegraphics[width=12cm]{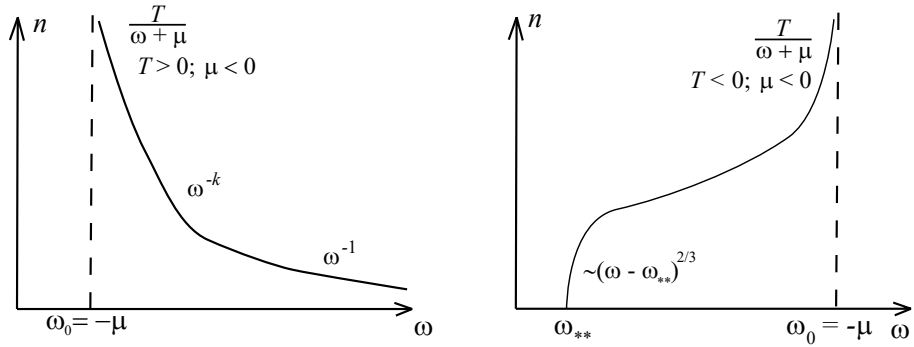}
\end{center}
\caption{ Sketches of spectra in case $k >1$ for orbits passing near $P2$.
{\bf Left:}   orbits from $O_{I}$.
{\bf Right:}  orbits from $O_{II}$.
}
\label{portrait4S}
\end{figure}

Correspondingly, the orbits from $O_{I}$  lying close to the separatrices $S_{IV,I}$ and $H$
correspond to  spectra $n(\omega)$ which start at a sharp front like in (\ref{S8c}), continue
with an intermediate KZ asymptotic and then asymptote to the thermodynamics spectrum~(\ref{S8e})
(see Figure~\ref{portrait4S}, left).
Note that presence of an intermediate KZ asymptotic is observed for all orbits
passing close enough to the fixed point $P2$ in the $O_{I}$, $O_{II}$, $O_{III}$ or $O_{IV}$ sectors.

The orbits from $O_{II}$ represent spectra vanishing at a finite point $\omega$
on the left boundary of the spectrum support.
They have an intermediate
KZ asymptotic (if the orbit pass close enough to $P2$).
Further, $n(\omega)$ blows up at a finite point $\omega$
on the right boundary of the spectrum support
(see Figure~\ref{portrait4S}, right).

The orbits from $O_{III}$ represent spectra vanishing at a finite points $\omega$
on both the left and the right boundaries of the spectrum support. Again, they have an intermediate
KZ asymptotic (if the orbit pass close enough to $P2$).

\begin{figure}[htbp]
\begin{center}
\includegraphics[width=12cm]{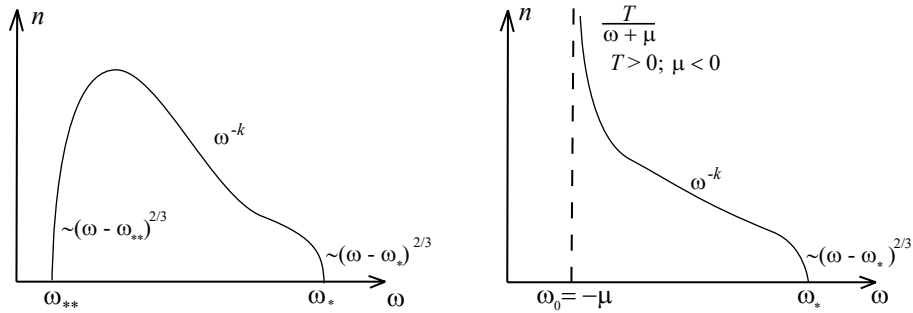}
\end{center}
\caption{ Sketches of spectra in case $k >1$ for orbits passing near $P2$.
{\bf Left:}  orbits from $O_{III}$.
{\bf Right:}  orbits from $O_{IV}$.
}
\label{portrait5S}
\end{figure}

The orbits from $O_{IV}$ which start on the $v$-axis, which corresponds to
a spectrum which has a left-boundary blowup of the form~(\ref{S8c}). Then the spectrum asymptotes to the KZ spectrum (if the orbit passes close enough to $P2$)  and finally $n(\omega)$ vanishes at a finite wave number
(see Figure~\ref{portrait5S}, right).

\newpage
The following two series of Figures demonstrate the behavior of spectrum near the separatrices with the KZ
scaling  on an intermediate range of frequencies.
The behaviour of spectrum passing far from $P2$ looks like on Figures~\ref{portrait4S}, \ref{portrait5S}
but without the intermediate KZ asymptotic.
\newpage

\section{Phase analysis for $k = 1$}\label{A}

The case $k = 1$ corresponds to the example of the direct cascade in the 2D NLS wave turbulence.
This case is degenerate since the KZ scaling $n \sim \omega^{-k}$
coincides with the thermodynamic scaling
$n \sim \omega^{-1}$.

The dynamical system reads
\begin{eqnarray}
&&\frac{du}{d\tau} = v, \label{A7}\\
&&\frac{dv}{d\tau} = -v + u^4 \label{A8}
\end{eqnarray}
and we have only a single fixed point $P1 = (0,0)$.
The differential matrix of the right-hand side of~(\ref{A7}, \ref{A8}) at $P1$ is
\begin{equation}\label{A9}
\Delta(0,0) =
\left(\begin{array}{cc}0&1\\ 0&-1\end{array}\right).
\end{equation}
Therefore $P1$ is a saddle-node with the eigenvalues $\lambda_1 = 0$, $\lambda_2 = -1$.
The stable manifold $S$ lies on the line $u+v=0$ while the central (slow) manifold  $U$ near $P1$ is
\begin{equation}\label{A10}
v = u^4;
\end{equation}
it expands from the origin into the half-plane $u > 0$.
Nontrivial dynamics of the dynamical system near by $P1$  can be described
 locally as s restriction of the phase flow to the central (slow) manifold, see~\cite{Carr}.
Namely, for small values of $\omega$,
near  $P1$, the stable manifold of $P1$ will dominate the dynamics and will quickly bring
the orbits very close to its slow  manifold unless the initial point is already on the slow manifold. The dynamics~(\ref{A7})
 in the vicinity of $P1$ on the slow manifold~(\ref{A10}) is  governed by
\begin{eqnarray}
\frac{du}{d\tau} = u^4, \label{A11}
\end{eqnarray}
which can be easily integrated,
\begin{equation}\label{A12}
u = \frac{1}{3^{1/3} (\tau_0 - \tau)^{1/3}}, \quad n = \frac{3^{1/3}\ln^{1/3}(\omega_0/\omega)}{\omega}.
\end{equation}
This is the log-corrected KZ spectrum. The thermodynamic spectrum in this case corresponds to the motion along the stable manifold giving $n\sim 1/\omega$.
\begin{lemma}\label{L2}
All orbits of the dynamical system~(\ref{A7}),~(\ref{A8})  starting in the first quadrant of the phase plane expand into the first quadrant never intersecting the $u$ and $v$-axes and without being bounded in either $u$ or $v$.
\end{lemma}
The latter assertion immediately follows from the observation that the vector field $(du/d\tau,dv/d\tau)$ restricted on these axes is directed into the first quadrant. Since there are no fixed points apart from $P1$,
by the Poincar\'e-Bendixon theorem it follows that the orbits must go to infinity in time $\tau$.
The orbits cannot be bounded in $u $: this follows from the integral identity
\begin{equation}\label{A13}
\frac{1}{2}u_{\tau}(\tau)^2 + \int_0^\tau u_s^2ds   = \Phi(u_0)  - \Phi(u(\tau)),
\end{equation}
where $\Phi(u) = - \frac{1}{5}u^5$, and the right-hand side of~(\ref{A13}) is always positive quantity.
If $u(\tau)$ was a bounded function for all $\tau$ then  $u_{\tau}(\tau)^2$ would also be a bounded quantity. Now suppose
$v$ is bounded and $u$ is unbounded. Then for  $u \to \infty$ one can neglect the first terms on the left-hand and the right-hand  sides of~(\ref{A13}). Differentiating the remaining equation, we get $u_\tau = u^4$, i.e. $v=u_\tau \to \infty$ which is a contradiction. Thus $v$ is unbounded and this completes the proof of
Lemma~\ref{L2}.

Consider the orbits from the fourth quadrant of the phase plane. Their behavior is analysed similar by the change of the time direction $\tau$ as $\hat\tau = - \tau$. With this, the orbits from the fist and fourth quadrants  (which are  the only ones physically admitted because $u$ is non-negative)
are divided into the three different classes: $O_I$, $O_{II}$ and $O_{III}$. The orbits $O_{I}$ and $O_{III}$ are separated by the
orbit $U$ which,  near $P1$,  is $v = u^4$ (the slow manifold). The  orbits $O_{II}$ of the fourth quadrant are the ones that intersect the $v$-axes for $v < 0$ with  slope  $-1$. The stable manifold separatrix $S$ separates the orbits $O_{II}$ and $O_{III}$. $S$
goes from infinity to  the equilibrium $P1$.
The orbits from $O_{III}$ are located between $U$ and $S$. These orbits go from infinity of the fourth quadrant always intersecting the $u$-axis and then approaching the separatrix $U$.

\begin{figure}[h]
\center{\includegraphics[width=8cm]{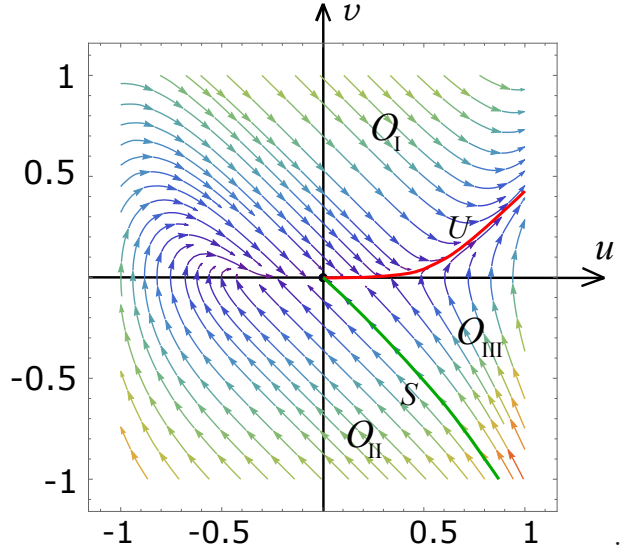}}.
\caption{Phase portrait of the dynamical system for $k = 1$.
The separatrices $U$ and $S$ are shown by different colour lines.
}
\label{portrait2A}
\end{figure}

Summarising, we have the following classification of the orbits, see Figure~\ref{portrait2A}.

\begin{theor}\label{T2}
$U$ and $S$ divide the phase half plane  $(0 < u < \infty,  -\infty < v < \infty)$ into parts $O_I$,  $O_{II}$ and  $O_{III}$ with the different behaviours of orbits. In part  $O_I$ (above $U$) the orbits intersecting the $u$-axis (never intersecting $v$-axis) asymptote to the separatrix $U$ as $\tau \to \infty$. In part $O_{II}$ (below $S$) the orbits
emerge out infinity of the fourth quadrant and go down along $S$ always intersecting the $v$-axis for $v < 0$.
Between $U$ and $S$, i.e. in $O_{III}$, the orbits go from infinity along $S$ emerging at the first quadrant and approaching $U$ as $\tau \to \infty$.
\end{theor}

Having this classification, we can proceed to the classification of solutions $n(\omega)$  as in section~\ref{S}.
Consider the orbits from $O_I$ and their dynamics near  the $v$-axis:
\begin{eqnarray}
&&\frac{du}{d\tau} = v, \label{A8a}\\
&&\frac{dv}{d\tau} = -v, \label{A8b}
\end{eqnarray}
Solving this system, we have
\begin{equation}\label{A8c}
u = v_0\left[e^{-\tau_0} - e^{-\tau}\right], \quad n = \frac{\omega_0}{v_0(\omega- \omega_0)}
\end{equation}
or in terms of the thermodynamical spectrum
\begin{equation}\label{A8d}
n = \frac{T}{(\mu + \omega)}.
\end{equation}
For $v > 0$ we have $ T > 0,  \mu < 0,  \omega > -\mu$ (left front blowup) and for
$v < 0$ we have $ T < 0,  \mu < 0,  \omega <- \mu$ (right front blowup).

The existence of central (slow) manifold leads the intermediate asymptotic
~(\ref{A12}) for orbits passing close enough to $P1$.

\begin{figure}[htbp]
\includegraphics[width=14cm]{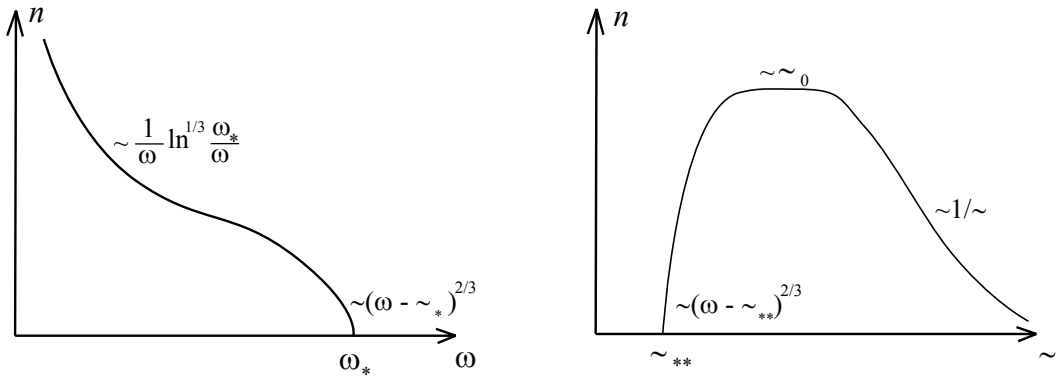}
\caption{
Sketches of spectra in the case $k = 1$.
{\bf Left}: the separatrix $U$.
{\bf Right}: the separatrix $S$.
}
\label{portrait3A}
\end{figure}

\begin{figure}[htbp]
\begin{center}
\includegraphics[width=16cm]{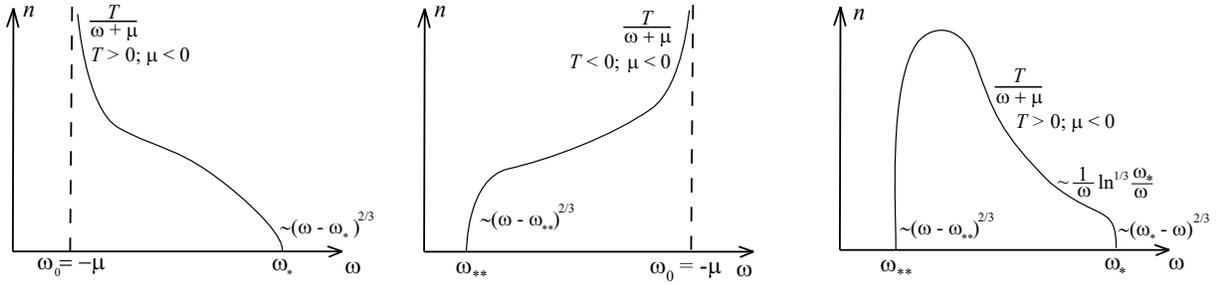}
\end{center}
\caption{
Sketches of spectra in the case $k = 1$.
{\bf Left}: $O_I$ orbits.  {\bf Centre}: $O_{II} $ orbits.
{\bf Right}: $O_{III}$ orbits.
}
\label{portrait4A}
\end{figure}

Finally as $\tau \to \infty$ i.e. for $u \gg 1$ and $v \gg 1$,  the dynamics is reduced to
\begin{eqnarray}
&&\frac{du}{d\tau} = v, \label{A7f}\\
&&\frac{dv}{d\tau} = u^4 \label{A7g}.
\end{eqnarray}
Again, $u(\tau)$ blows up at a finite time $\tau_*$.
Respectively,  spectrum $n(\omega)$ vanishes at $\omega_* < \infty$,
\begin{equation}\label{A8k}
n(\omega) \sim (\omega - \omega_*)^{2/3}
\end{equation}
as $\omega \to \omega_*$ and $\omega < \omega_*$. Similarly for for $u \gg 1$ and $-v \gg 1$, using the the inverse time direction
$\hat\tau$, we get that  spectrum $n(\omega)$ vanishes at $\omega_{**} < \infty$,
\begin{equation}\label{A8k}
n(\omega) \sim (\omega_{**} - \omega)^{2/3}.
\end{equation}
The behaviour of  spectrum $n(\omega)$ described above is summarised in Figures~\ref{portrait3A} and~\ref{portrait4A}.


\newpage

\section{Phase analysis for $0 < k < 1$}\label{B}

The case $0 < k < 1$  is realised for the 3D NLS inverse cascade  ($k=7/6$), and
the SNE turbulence in all the cases -- the 2D direct cascade ($k=1/3$), the 3D direct cascade  ($k=5/6$),
the 2D inverse cascade ($k=0$) and the 3D inverse cascade  ($k=1/2$).

This is the case when the pure KZ spectrum cannot be realised because it formally corresponds to a negative
spectrum, $n = (k(k-1))^{-1/3}\omega^{-k} <0$\footnote{Alternatively, one could say that the spectrum is positive but the flux is in a ``wrong" direction, ${\cal P} < 0$. This becomes evident if we rescale our solution to consider cases with arbitrary values of  ${\cal B} =  {\cal P}/{\cal I} $, which gives $n = (k(k-1)   {\cal P}/{\cal I} )^{-1/3}\omega^{-k} $. }

 With $k < 1$ we have only a single fixed point of the dynamical system~(\ref{S7}), (\ref{S8})
 in the right half plane of the phase plot, $u\ge 0$:  $P1 = (0,0)$.
  The eigenvalues of the linearised
matrix at $P1$ equal $\lambda_1 = - k < 0$ and $\lambda_2 = - k + 1>0$, i.e. $P1$ is the saddle point.
The eigenvectors are $\xi_1 = (1,\lambda_1)$ and $\xi_2 = (1,\lambda_2)$.
\begin{lemma}\label{L3}
All orbits of the dynamical system~(\ref{S7}),~(\ref{S8})  starting in the first quadrant of the phase plane expand into the first quadrant never intersecting the $u$ and $v$-axes and without being bounded in either $u$ or $v$.
\end{lemma}
The proof of this Lemma completely repeats the proving Lemma~\ref{L2} --
now using the integral identity~(\ref{S12}). Again similarly, we analyse the fourth quadrant.

\begin{figure}[h]
\center{\includegraphics[width=12cm]{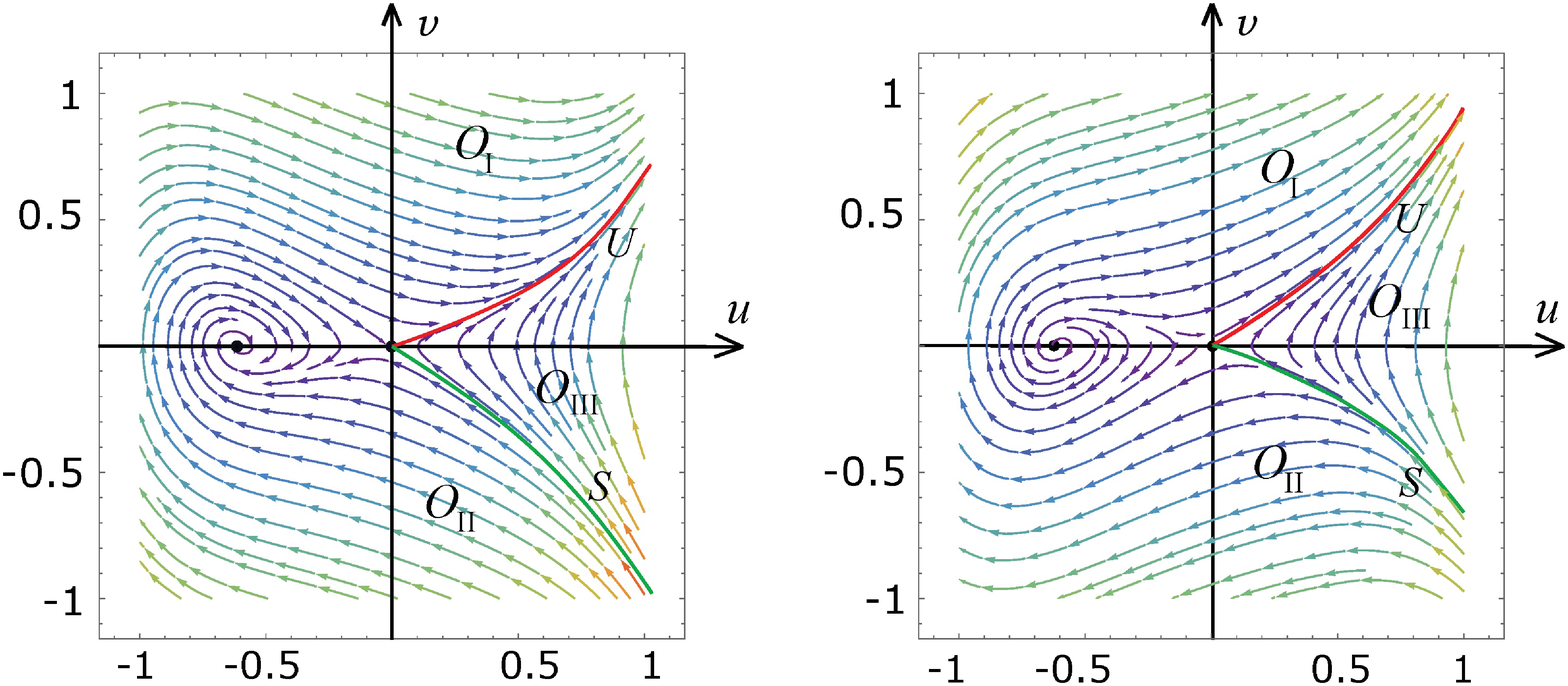}}.
\caption{Phase portrait of the dynamical system for $k = 1/3$ (the left panel) and $k = 2/3$ (the right panel).
The separatrices are shown by different colour lines.
Note that in spite the different stability properties of the fixed point $P2$ on the left half-plane, the qualitative properties of the right half-plane orbits are the same.
}
\label{portrait1B}
\end{figure}

Thus, similar to how it was done in section~\ref{A}, the orbits from the first and fourth quadrants
are divided into  three different classes: $O_I$, $O_{II}$ and $O_{III}$.  Let us denote the stable and unstable manifolds
of $P1$ by $S$ and $U$. The orbits $O_{I}$ and $O_{III}$ are separated by the
orbit $U$,
whereas the orbit $S$  separates the orbits $O_{II}$ and $O_{III}$.
With this, we get the following classification of the orbits, see Figure~\ref{portrait1B}.
\begin{theor}\label{T3}
The stable and unstable manifolds of $P1$,  $S$ and $U$, divide the phase plane to $O_I$, $O_{II}$ and $O_{III}$ parts with
different behaviour of orbits. The orbits from $O_I$ (above $U$) approach the separatrix $U$  with
$(u,v) \to (\infty,\infty$) as $\tau$ evolves. Between $U$ and $S$, i.e. in $O_{II}$, the orbits arrive
from infinity along $S$ towards $P1$ intersecting the $u$-axes and then turn back asymptoting to
the separatrix $U$ in $\tau$. In part $O_{III}$ (below $S$) the orbits emerge out infinity in the fourth
quadrant always intersecting the $v$-axis for $v < 0$.
\end{theor}

\begin{figure}[htbp]
\begin{center}
\includegraphics[width=12cm]{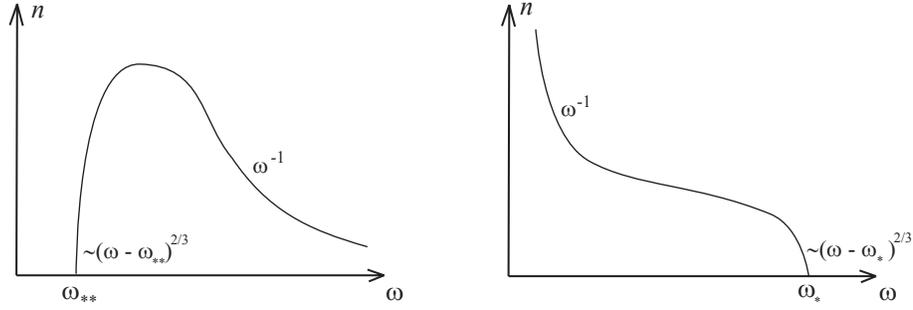}
\end{center}
\caption{ Sketches of spectra in the case $k = 2/3$.
{\bf Left:} separatrix $S$.
{\bf Right:} separatrix $U$.
}
\label{portrait2B}
\end{figure}

Similarly to how it was done in the case $k > 1$ the dynamics
of the separatrix $U$ as $\tau \to \infty$ in the leading order is governed by
the dynamical system~(\ref{S7a}, \ref{S7b}).
It means that
\begin{equation}\label{BS12a}
u(\tau) \sim (\tau_* - \tau)^{-2/3}, \quad \tau < \tau_* < \infty.
\end{equation}
The spectrum $n(\omega)$ behaves as
\begin{equation}\label{BS12b}
n(\omega) \sim ({\omega_*}   - {\omega})^{2/3} \quad {\rm as } \quad \omega \to \omega_* < \infty.
\end{equation}
The separatrix $S$ for $u \gg 1$ and $v \ll - 1$ is analysed similarly by changing
the direction of time $\tau$ as
$\hat\tau$. As a result, we get
\begin{equation}\label{BS13}
u(\hat\tau) \sim (\hat\tau_{**} - \hat\tau)^{-2/3}, \quad \hat\tau < \hat\tau_{**} < \infty.
\end{equation}
Then the spectrum $n(\omega)$ behaves as
\begin{equation}\label{BS14}
n(\omega) \sim
 ({\omega}   - {\omega}_{**})^{2/3} \quad {\rm as } \quad \omega \to \omega_{**} < \infty.
\end{equation}
The corresponding spectra are presented on Figure~\ref{portrait2B}.

Behaviour of the orbits near the $v$-axis, $u \ll v $, is again derived from the system~(\ref{S8a}, \ref{S8b})
i.e. using the formula~(\ref{S8c})\footnote{Note that formula~(\ref{S8c}) was obtained for $k \ne 1/2$, but it is easy to see that the same asymptotic formula appears in the case $k=1/2$ too.}
\begin{equation}\label{BS15}
n  =  \frac{T}{\mu + \omega},
\end{equation}
where $T=\omega_0^{ k}/v_0  $ and $\mu = - \omega_0 <0$.
For the orbits from $O_{I}$ we have $T>0$ (left boundary blowup) and
for the orbits from $O_{II}$ we have, respectively,  $T<0$
 (right boundary blowup).

The behaviour of spectrum $n(\omega)$ for $u, v \ll 1$~i.e. near  $P1$ is
\begin{equation}\label{BS16}
n  \sim \omega^{-1}.
\end{equation}

\begin{figure}[htbp]
\begin{center}
\includegraphics[width=16cm]{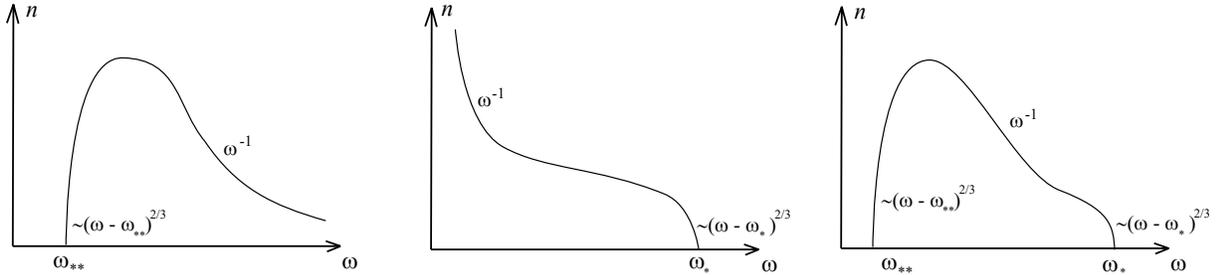}
\end{center}
\caption{ Sketches of spectra in case $k = 2/3$.
{\bf Left:} spectrum behaviour for the orbits $O_{I}$.
{\bf Centre:} spectrum behaviour for the orbits $O_{II}$.
{\bf Right:} spectrum behaviour for the orbits $O_{III}$.
}
\label{portrait3B}
\end{figure}

Summarising, the orbits $O_{I}$ start on the $v$-axis with
the spectrum which has the form~(\ref{BS15}). Then they asymptote to the separatrix $U$ as $\tau $ evolves and finally $n(\omega)$ vanishes at a finite frequency; see the left panel of Figure~\ref{portrait3B}. The orbits form $O_{II}$ has the spectrum behaviour in the reverse order with respect to $O_{I}$; see the centre panel of Figure~\ref{portrait3B}.
Finally, orbits from $O_{III}$ correspond to spectra that are always compactly supported
~i.e. $n(\omega) > 0$ only on a bounded interval $(\omega_*,\omega_{**})$, see the right panel of Figure~\ref{portrait3B}.
They intermediately asymptote to the pure thermodynamical spectrum $\omega^{-1}$ if their orbits pass near  the equilibrium $P1$. Note that passing close to $P1$ means that the forcing and the dissipation scales  are well separated (since the motion is very slow near $P1$).

\section{Discussion}\label{C}
In this paper, we have presented an exhaustive study and  a full classification  of all possible one-flux steady states  for both the direct and inverse cascade solutions of
the differential kinetic equation of local wave turbulence~(\ref{S6}).
The behaviour is shown to be qualitatively different in three cases: $k>1$, $k=1$ and $k<1$ where $k=({s-2})/{3}$ is the KZ exponent.

In particular, the KZ scaling can be observed (in its pure form or as an asymptotic) only for $k>1$.
``Warm" cascades realising states with mixed thermodynamic and flux components are also possible in this case.
They are represented by a great variety of spectra with different values of thermodynamic potentials $T$ and $\mu$, some
of them characterised by a sharp front or/and a blowup near the forcing or/and dissipation scale.
The spectrum blowup behaviour could be interpreted as a condensation phenomenon.
Which of the spectrum is realised is determined by the type of forcing and dissipation in each particular case.

Case $k=1$ is degenerate: the KZ and the thermodynamic exponents coincide. Thus, a log-corrected spectrum is realised instead of the pure power-law KZ spectrum.
No KZ scaling can be realised in the case $k<1$, neither in its pure form nor as an asymptotic. All spectra are ``warm" in this case: they are close to the pure thermodynamic Rayleigh-Jeans spectrum deeply in the inertial range if the forcing and the dissipative scales are well-separated.

Finally, for any $k$ the solutions are divided into  distinct classes corresponding to the orbits of the phase plane divided by separatrices which connect fixed points of the corresponding dynamical system with each other or with infinity.
There are four of such classes in the case $k>1$ ($O_{I}$, $O_{II}$, $O_{III}$ and $O_{IV}$) and three classes for $k\le1$
($O_{I}$, $O_{II}$, $O_{III}$).

As a concluding remark we note that, in spite of a great variety of the found solutions, there exist also stationary solutions in which both fluxes $P$ and $Q$ are present simultaneously. These solutions are important in cases where forcing is present at two or more different scales. Unfortunately in such cases the respective dynamical system is not autonomous and cannot be analysed by the phase-plane method presented in this paper. However, it would be interesting in future to analyse such solutions, for instance numerically.

\section*{Acknowledgements}

Sergey Nazarenko is supported by the Chaire  D'Excellence IDEX (Initiative of Excellence) awarded by
Universit\'e de la C\^ote d'Azur, France,   Simons  Foundation Collaboration grant Wave Turbulence (Award ID 651471), the  European  Unions  Horizon  2020
research and innovation programme  in the framework of Marie Skodowska-Curie HALT project (grant agreement No 823937) and  the FET Flagships PhoQuS project
(grant agreement No 820392).
Vladimir Grebenev's, Boris Semisalov's and work on this project was partially supported by the ``chercheurs invit\'es" awards of
the F\'ed\'eration Doeblin FR 2800, Universit\'e de la C\^ote d'Azur, France.
Sergey Medvedev's and work on this project was partially supported by CNRS
``International Visiting Researcher" award.

\end{document}